\documentclass[prl,twocolumn,epsf,psfig]{revtex4}
\usepackage{amssymb}

\def\widetext{\end{twocolumn}}
\input epsf
\begin{document}

\title{Dynamically generated dimension reduction and crossover in a 
spin orbital
model}

\author{Theja N. De Silva$^{a,b}$, Michael Ma$^{a,c}$, and
Fu Chun Zhang$^{a,d}$}
\affiliation{$^{a}$Department of Physics, University of Cincinnati,
Cincinnati Ohio 45221 \\
$^{b}$Department of physics, University of Ruhuna, Matara, Sri Lanka  \\
$^{c}$ Department of Physics, Hong Kong University of Science and Technology\\
$^{d}$ Department of physics, University of Hong Kong, Hong Kong}

\begin{abstract}
\noindent  
We study a spin orbital model in which the spin-spin interaction couples linearly to
 the orbital isospin. Fluctuations drive 
the transition from paramagnetic state to C type ordered state into a strongly first 
order one, as observed
in $V_2O_3$. At $T=0$, there is a ferro-orbital-C-spin to ferro-orbital-G-spin transition. Close to the 
transition point, the system shows dynamically generated dimension reduction 
and crossover, resulting in one or more spin reentrant transitions. 
\end{abstract}

\maketitle

Recently there has been growing interest in the effects of orbital
degeneracy on the physics of transition metal oxides. The effective spin
Hamiltonian of the insulating phase of such systems may depend crucially on
orbital short-ranged and long-ranged correlations. As a result, magnetic
ordering can become anomalous or may even be suppressed altogether. At the
same time, orbital physics is also affected by spin fluctuations and
correlations. The interplay between spin and orbital degrees of freedom is
fundamental to much of the physics of transition metal oxides~\cite
{tokura,imada}. In this paper we investigate certain aspects of this
interplay, emphasizing on the effects of thermal and quantum spin
fluctuations on orbital ordering and the effects of orbital ordering on spin
physics. Our most interesting result is that the system can exhibit dimension reduction and
dimension crossover of spin physics as a function of temperature due to orbital ordering.

In the insulating phase of transition metal oxides, the dominating energy
scales for the transition metal ions are the on-site Coulomb repulsion,
Hund's rule coupling, and the crystal field due to the surrounding oxygen
ions. Neglecting weak
spin-orbit effect, the general spin Hamiltonian with two-fold degenerate
orbitals (represented by pseudospin $\tau =1/2$) is of the form~\cite{kugel}
, $H=\sum_{\left\langle ij\right\rangle }(J_{ij}\mathbf{S}_{i}\cdot \mathbf{S}%
_{j}+K_{ij})$, where $J_{ij}$ and $K_{ij}$ are functions of $\tau _{i}$, and $%
\tau _{j}$. This Hamiltonian has global $SU(2)$ invariance in spin space and
a lower and discrete rotational symmetry in $\tau $ space~\cite
{khomskii,castellani,joshi,theja,kha}. Out of the general class of such
Hamiltonians, we will focus on those on a cubic lattice of the form
\begin{eqnarray}
H=J_{0}\sum_{\left\langle ij\right\rangle }\mathbf{S}_{i}\cdot \mathbf{S}
_{j}-K\ \sum_{\left\langle ij\right\rangle }\mathbf{S}_{i}\cdot \mathbf{S}%
_{j}(\mathbf{\tau }_{i}\cdot \widehat{\mathbf{n}}_{ij}+\mathbf{\tau }
_{j}\cdot \widehat{\mathbf{n}}_{ij})
\end{eqnarray}
\noindent Here the unit vectors $\widehat{\mathbf{n}}_{ij}=\widehat{\mathbf{n%
}}_{1},$ $\widehat{\mathbf{n}}_{2},$ $\widehat{\mathbf{n}}_{3}$ for $i,j$
nearest neighbor in the $x,$ $y,$ $z$ directions respectively. In this
model, the interplay between spins and orbitals arises from the second term
which is linear in $\tau $. This linear term will be present provided the
two eigenvalues of the hopping matrix are different, while the $\widehat{%
\mathbf{n}}_{i}$'s will depend on how the two degenerate orbitals transform
under lattice rotations. For specificity, we take the $\widehat{\mathbf{n}}%
_{i}$ 's to be unit vectors in the $x-z$ plane, with $\widehat{\mathbf{n}}%
_{3}=\widehat{z},$ while $\widehat{\mathbf{n}}_{1}$ and $\widehat{\mathbf{n}}%
_{2}$ are rotated from $\widehat{\mathbf{n}}_{3}$ by 120$^{0}$ and 240$^{0}.$
More generally, there are also quadratic in $\tau $ terms, which we assume
to be weak compared to the two terms kept. This implies that any orbital
ordering in the system will be due to spin-orbital coupling rather than
Jahn-Teller effect. Assuming this is the case, and with the choice of $%
\widehat{\mathbf{n}}_{i}$ 's above, this spin-orbital Hamiltonian can serve
as a model for one electron or hole per site in the doubly degenerate $e_{g}$
levels of cubic peroskites~\cite{khomskii} as well as a possible model for $%
V_{2}O_{3}$~\cite{mila}. For the latter, each site on the cubic lattice is
the topological equivalence of a vertical pair of sites on the corundum
lattice of $V_{2}O_{3}$. The coupling $J_{0}$ depends strongly on and
decreases with the Hund's coupling, while $K$ is only weakly dependent on
it. We consider $J_{0}>0$, and with an
appropriate definition of $\tau $, we also have $K>0$. The calculation shown
here will be for $S=2$, the value of $S$ for the $V_{2}O_{3}$ bond model,
but the results are qualitatively the same for other $S$. The results are
also applicable to other lattices and other choices of the $\widehat{\mathbf{%
n}}_{i}$'s. 

While $J_{0}$ favors conventional (G-type) AF magnetic correlations
so that nearest neighbor spins are all AF correlated,
 $K$ favors, along with orbital ordering, anomalous magnetic
correlations break the cubic lattice rotational symmetry,
for example C-type ordering with AF correlation in ab plane and FM correlation in c direction. 
In this paper, we investigate the
phase diagram of this model in the temperature $T$ and $J_{0}/K$ plane. Our
main results are: 1) At low $T,$ orbital ordering gives rise to effective
dimension reduction of the spin physics for $J_{0}/K$ close to $2.$ 2) The
weakening of orbital ordering with increasing $T$ can lead to dimension
crossover from $2D$ to $3D$ and vice versa. 3) The dimension crossover
effect together with thermal fluctuation effects on the spins can lead to an
order by disorder mechanism and one or more reentrant transitions. 4) The
strongly first order nature of the magnetic transition in $V_{2}O_{3}$ is
explained~\cite{bao2}. The underlying physics behind these results are orbital ordering
coupled with spin short-ranged correlations and quantum fluctuations.

Within the context of the bond model for $V_{2}O_{3},$ our model has been
studied by Joshi \textit{et. al. }using a single-site mean field theory. In
order to include short range spin correlation and quantum fluctuations, we
use a modified mean field approach to decouple the spin orbital
Hamiltonian. We begin with the Feynman-Hellman theorem~\cite{fe}:
\begin{eqnarray}
F\le F_0+\left\langle H-H_0\right\rangle_0
\end{eqnarray}
Here $F$ is the true free energy of the system and $H$ is the actual
Hamiltonian given in Eq. 1. $H_{0}$ is a variational Hamiltonian and $%
\left\langle {}\right\rangle _{0}$ is the thermal expectation value with
respect to $H_{0}$. $F_{0}$ is the free energy of the system with
Hamiltonian $H_{0}$. We take our variational Hamiltonian as,
\begin{eqnarray}
H_0=H_S+H_{\tau} \\
H_S=J_{\perp}\sum_{\left\langle ij\right\rangle }^{z} \mathbf{S} _{i}\cdot
\mathbf{S}_{j}+J_{\parallel}\sum_{\left\langle ij\right\rangle }^{x,y}
\mathbf{S}_{i}\cdot \mathbf{S}_{j}  \nonumber \\
H_{\tau}=-2KA\sum_{i}{\tau _{iz}}  \nonumber
\end{eqnarray}
\noindent where, $J_{\perp }=J_{0}-2Kt$, and $J_{\parallel }=J_{0}+Kt$, with
${t\ge 0}$. In $H_{S}$ above, the first sum is for nearest neighbor pairs
along the z direction and the second sum is for pairs on the same xy plane. $%
t$ and $A$ are variational parameters. Our choice of $H_{0}$ is based on
the expectation that orbital ordering will be ferro-orbital. 

Minimization of Feynman-Hellman free 
energy with respect to the variational parameters gives two self consistent equations, 
$A=\Delta{B}$ and $t=\left\langle \tau _{z}\right\rangle=\tanh (2K\beta A)$. 
$\Delta{B}=B_{\perp }-B_{\parallel}$, with
$B_{\perp } =\left\langle {\bf S}_{i}\cdot {\bf S}_{j}\right\rangle_{z}$ and $B_{\parallel}= \left\langle {\bf S}_{i}\cdot {\bf S}_{j}\right\rangle_{x,y}$ as out of plane and in-plane nearest neighbor spin spin correlation obtained from $H_S$.

Note that while non zero
value of $t$ signifies long range orbital order, non zero value of $B$ only
signifies short range spin correlations. If $B_{\perp }\neq B_{\parallel },$
spin correlations will be different from isotropic G type antiferromagnet.
In particular, if $B_{\parallel }<0,$ and $B_{\perp }>0,$ that would
correspond to C type magnetic correlations. Our decoupling scheme allows us
to study the effects of short-ranged spin-spin correlations, but not
short-ranged orbital correlations. However, we expect spin fluctuations to
be dominant because of its continuous symmetry. Joint spin-orbital
correlations are also ignored, but we expect these to be weak compared to
spin fluctuations far away from the $SU(4)$ limit~\cite{li}.

Note that $J_{\parallel }>0$, but $J_{\perp }$ can be either positive or
negative in $H_{S}$, depending on the value of orbital order parameter $t$. $%
H_{S}$ is an Heisenberg Hamiltonian with spatial anisotropy. In order to
include short-ranged correlations and quantum fluctuations, we use
renormalized spin wave theory (RSWT)~\cite{takahashi,hirsch,tang}, which unlike
traditional spin wave theory (SWT) is applicable to both magnetically
ordered and disordered phase. For the Heisenberg antiferromagnet on the
square lattice, Hirsch and Tang~\cite{hirsch} have shown that this method
can provide quantitatively accurate results. In RSWT, magnon-magnon
interactions are approximated by introducing a constraint that the total
staggered magnetization be zero; $M=\sum_{i\in A}S_{i}^{z}-\sum_{j\in
B}S_{j}^{z}=0$, or equivalently that the average number of spin waves per
site is $S$. This constraint can be implemented by introducing into $H_{S}$
a Lagrange multiplier $\lambda $:
\begin{eqnarray}
H_S =J_{\perp}\sum_{\left\langle ij\right\rangle }^{z} \mathbf{S}_{i}\cdot
\mathbf{S}_{j}+J_{\parallel}\sum_{\left\langle ij\right\rangle }^{x,y}
\mathbf{S}_{i}\cdot \mathbf{S}_{j}- \lambda M
\end{eqnarray}
\noindent The modified $H_{S}$ (Eq. 4) is then solved using usual spin wave
theory by expanding to quadratic order in Holstein-Primakoff bosons.

Classically, the spins will order as C-type and G-type for the $J_{\perp }<0$
and $J_{\perp }>0$ cases respectively. Thus, the sublattice designation will
differ in the two cases and the spin wave calculation must be done
separately. Let us define $Q=\frac{J_{\perp }}{J_{\parallel }}$ and $\omega
_{k}^{+}$ and $\omega _{k}^{-}$, magnon energies for $Q>0$ and $Q<0$ cases
respectively. We see that $Q$ is a dimensionless measure of the effective spin-spin coupling anisotropy. Note that $-2<Q<1$. 
The chemical potential $\mu $ is obtained from the constraint equation,
\begin{eqnarray}
S+ \frac{1}{2}= \int\frac{d^{d}\vec k} {(2\pi)^d} \biggl(\frac{1}{e^{\beta
\omega_k^{\pm}}-1}+ \frac{1}{2}\biggr)F_{\pm}(\mu,k),
\end{eqnarray}
\noindent where $\beta $ is the inverse temperature and $\omega _{k}^{\pm }$
are given by, $\omega _{k}^{+}=6J_{\parallel }S\sqrt{{\mu }^{2}-{\gamma _{k}}%
^{2}}$ and $\omega _{k}^{-}=6J_{\parallel }S\sqrt{(\mu -\frac{|Q|}{3}\gamma
_{k\perp })^{2}-(\frac{2}{3}\gamma _{k\parallel })^{2}}$ with effective
magnon chemical potential, $\mu ={\frac{1}{3}}(2+|Q|+\frac{\lambda }{%
4SJ_{\parallel }})$. $\gamma _{k}=\frac{1}{3}(\cos {k_{x}}+\cos {k_{y}}%
+Q\cos {k_{z}})$, $\gamma _{k\perp }=\cos {k_{z}}$ and $\gamma _{k\parallel
}=\frac{1}{2}(\cos {k_{x}}+\cos {k_{y}})$. $F_{+}(\mu ,k)=\mu (\mu ^{2}-\gamma _{\vec{k}}^{2})^{-1/2}$
and $F_{-}(\mu ,k)=(\mu +\frac{Q}{3}\gamma _{k\perp })((\mu +\frac{Q}{3}%
\gamma _{k\perp })^{2}-(\frac{2}{3}\gamma _{\vec{k}\parallel })^{2})^{-1/2}$
should be used for $Q>0,$ $<0$ respectively.

After solving for $\mu$, $\Delta{B}$ can 
be calculated, and the self consistent equation for $t$ can be solved using iterative scheme. 
When there are more than one solutions for $t,$ we compare their free
energies to choose the stable solution for each temperature.  One anomaly of
the RSWT approach is that  $B_{\perp }$ and $B_{\parallel }$ need to be
calculated to one order of $1/S$ higher than the free energy to ensure the
correct sign for spin correlation~at high temperatures\cite
{takahashi,takahashi2}. This difference is not significant for the
self-consistent solutions close to and below the temperature of the first
phase transition in our model.

In FIG. 3, we show the resulting phase diagram for the case of $S=2,$ the
value of $S$ for the $V_{2}O_{3}$ bond model. In what follows, P stands for
para (i.e. disordered), F stands for ferro ordering, C stands for C type
ordering, G stands for G-type ordering, and O and S refer to orbital and
spin respectively. Five phases are possible in the model. These phases are
POPS , FOCS, FOPS (may be GS or CS short-ranged correlations), POGS (isotropic
AFLRO), FOGS (anisotropic AFLRO). The phase consistent with the magnetic and
orbital ordering observed in \ $V_{2}O_{3}$ is the FOCS phase \cite{mila,bao2,pao,park}. The phase
transitions the system undergoes as the temperature
is lowered depend on the parameter $J_{0}/K$ and can be grouped into 6 regimes discussed below. In all cases, the orbital transition
is first order while the spin transitions are second order unless
accompanied by the orbital transition.
\begin{figure}[tbh]
\epsfxsize=7.1cm 
\centerline{\epsffile{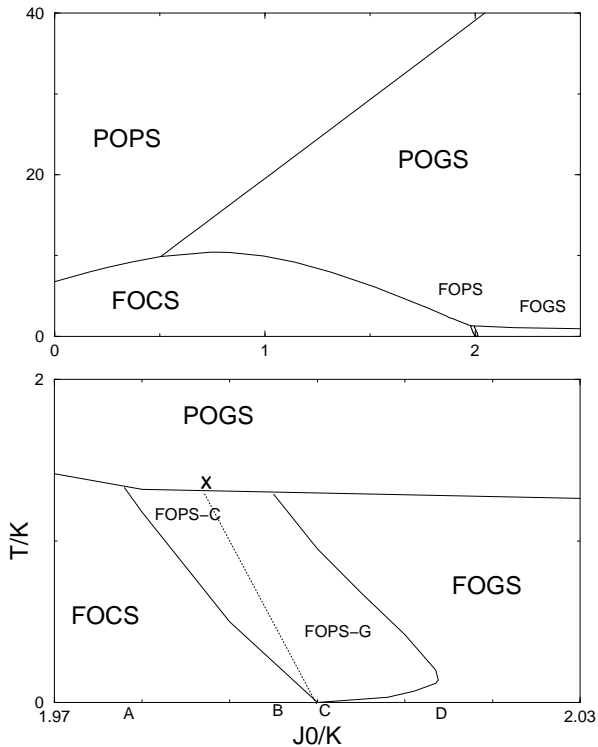}}
\caption{ (a) Phase diagram of spin-2 doubly degenerated spin orbital
model. (b) Blow up of the region close to $J_{0}/K=2.$ Regimes III, IV, and V
described in the text correspond to values of $J_{0}/K$ between
AB, BC, and CD respectively. Dotted line gives $T_{2},$ the dimension reduction
temperature. To its left (right), the spin coupling anisotropy parameter $Q<0 (>0)$.}
\end{figure}

For regime I ($0 \le \frac{J_0}{K} \le 0.506$), the regime relevant for $V_{2}O_{3},$ the
transition is a direct one from the disordered phase into the FOCS phase
with both spins and orbitals ordered. Experiments showed this transition to be strongly
first order. While single site mean field theory (SSMFT) for our Hamiltonian
is able to obtain a first order transition in this parameter regime, it was
only weakly first order. By including short-ranged correlations in the
present theory, the POPS phase is stabilized, and the first order nature of
the transition is significantly enhanced. Entropy jump calculated from the free 
energy derivative, is larger than $1k_{B} $ per site throughout regime I, 
with e.g. a value of $1.28k_{B}$ for
$J_{0}/K=0.25$, which compares favorably with the experimental value of 
$1.4k_{B}$ (each site on the bond model corresponds to a pair of V ions). We
should note that the
mechanism for large entropy jump here is quite different from the usual fluctuation driven first
order transition where the dominant fluctuations are from modes close to the
mean field free energy minimum. For this system, however, the dominant
fluctuations are G-type magnetic ones, and very far from the C-type magnetic
ordering. Another issue in $V_{2}O_{3}$ is why spins and orbitals order at
the same temperature. In principle, it is possible to have a spin Peierls transition
driven by FO ordering instead of a FOCS transition~\cite{joshi}. This issue cannot be
addressed by SSMFT, but can be using our modified MFT, which shows that for $%
J_{0}/K$ appropriate for $V_{2}O_{3},$ there is no FO driven spin Peierls
phase. However, we will see below that such a phase can indeed occur in
other parameter regimes of our model, but it will necessarily be preceded by
a POGS ordering at higher temperature. Thus, a phenomenological explanation
of the concurrence of orbital and CS ordering in $V_{2}O_{3}$ is that there
is no GS ordering at a higher temperature.

Regimes II and VI ($0.506 < \frac{J_0}{K} \le 1.976$ and $2.013 < \frac{J_0}{K}$ respectively) show two phase transitions. The system first undergoes an
isotropic ($Q=1$) POGS ordering. Then at a lower temperature, the
spin-orbital coupling causes a first order FO transition that converts the
spin ordering to CS ($Q<0$) in regime II and aniostropic GS ($Q>0,$ $\neq 1)$
in regimes VI. Rather more interesting, however, are regimes III, IV, and V, corresponding to $1.976 < \frac{J_0}{K} \le 1.987$,  $1.987 < \frac{J_0}{K} \le 2$,  and $2 < \frac{J_0}{K} \le 2.013$ respectively. These regimes show multiple transitions, including reentrance. These transitions are
consequences of effective dimension reduction and dimension crossover in
spin physics caused by orbital ordering. 

To see this, let us first
consider $T=0,$ where within our MFT the orbital is always fully ordered. As
a result, the effective spin Hamiltonian from our decoupling scheme
parameterized by $Q$ changes \textit{continuously} to smaller positive $Q$
and then eventually to negative $Q$ as $J_{0}/K$ is decreased. At $%
J_{0}/K=2, $ $Q=0$ and the spins on different planes become decoupled, i.e.
the spin Hamiltonian is that of a 2D Heisenberg antiferromagnet. This is the
orbital driven dimension reduction effect at $T=0.$ Current wisdom is that
the ground state of the 2D Heisenberg anitferromagnetic Hamiltonian is
ordered at $T=0$ in 2D even for $S=1/2.$ Thus there is spin LRO (anisotropic
GS or CS) for all values of $J_{0}/K$ at $T=0.$ At finite temperature,
thermal fluctuations will weaken both the spin ordering and the orbital
ordering, with the latter giving rise to effective temperature-dependent
spin Hamiltonian. The spin physics is best understood by considering how $%
Q(t)$ changes with $T$ together with the dependence of the spin transition
temperature $T_{c}(Q)$ on $Q.$ For small $\left| Q\right| ,T_{c}(Q)\sim
\left| Q\right| ^{\theta }.$ RSWT gives $\theta
=1/2$. Because orbital has a
discrete symmetry, its order parameter $t$, and hence $Q$ changes
exponentially slowly at low $T,$ and the physics is dominated by thermal
disordering of the spin. At higher $T,$ the reduction in $t$ becomes
significant, and the corresponding change in $Q$ can give rise to dimension
crossover in spin physics. These features are shown in Fig.3b and discussed 
 below, where $Q_{0}$ denotes the value of $Q$ for $t=1,$ i.e. at $%
T=0.$

We first consider right at the decoupling point $J_{0}/K=2$ (point C in Fig.
3b), so that $Q_{0}=0$ at $T=0.$ As $T$ increases, $t$ decreases and $Q$
becomes increasingly positive, the planes become increasingly coupled,
implying a crossover from 2D to 3D. However, this crossover is exponentially
slow at low $T,$ and since Heisenberg spins cannot order at any $T>0$ in 2D,
the spin LRO is immediately destroyed at infinitesimal $T$. As temperature
increases, $Q$ becomes large enough that $T_{c}(Q)$ exceeds $T$, and there
is a reentrant transition into an anisotropic GS phase. The restoration of
spin LRO due to temperature induced dimension crossover can be viewed as a
novel kind of order by disorder mechanism. For $Q_{0}>0$ but small (regime
V), the physics is basically the same with one difference. Since now $%
T_{c}(Q_{0})>0,$ the GS order is stable at low $T$ but will disorder for $%
T\gtrsim T_{c}(Q_{0}).$

The behavior is even richer for $Q_{0}<0$ but small (regimes III and IV).
Now as $T$ increases, $Q$ gets first less negative, becomes $0$ at some
temperature $T_{2},$ then becomes positive. That is we have dimension
crossovers first from anisotropic 3D to 2D and then back to anisotropic 3D
as $T$ increases. Correspondingly, the spins first undergoes a transition
from CS LRO to CS short ranged order at $T\approx T_{c}(Q_{0}).$ The
interplane ferrromagnetic correlation continues to decrease as $T$
increases, crossing over into GS short-ranged order for $T>T_{2}.$ In regime
IV, there is yet another reentrant transition into anisotropic GS LRO.
Throughout regimes III - V, the PS phases has short-ranged spin correlations
that are spatially anisotropic and so break the lattice rotational
symmetry. In effect, these are orbital driven spin Peiriels phases.
Eventually, in all these regimes, the system switches back in a first order
jump back into isotropic GS ordering when the orbital becomes disordered.
Point X is where this transition coincides with the dimension reduction
temperature $T_{2},$ so amazingly there is a jump directly from isotropic 3D
behavior just above the transition to exactly 2D just below.

In summary, we have investigated the problem of the interplay between spins
and orbitals in transition metal oxides concentrating on the competition
between spin-spin interactions and spin-orbital coupling. In addition to
illuminating the phase transition properties of magnetic ordering in
insulating $V_{2}O_{3},$ our model shows a mechanism for dynamically
generated dimensional reduction and dimension crossover. Although the
results presented are for $S=2,$ the same qualitative behavior will hold for
other $S.$ Also, while our calculations are restricted to the Hamiltonian (Eq. 1), 
these effects will be present in other spin-orbital models as long as orbital 
ordering results in vanishing spin-spin coupling in one or more spatial directions. 
Since these dimensional reduction and crossover effects are
present only close to the decoupling point, to observe them one would need
to find systems with the appropriate Hund's coupling so as to produce the
proper $J_{0}/K$ range. For $S=2,$ there is the additional problem that this
range is very narrow. Larger range will occur for smaller $S$. 
Therefore it will be interesting to search for
effective $S=1/2$ peroskite transition metal oxides with double orbital
degeneracy and weak Jahn-Teller coupling.

This work was in part supported by DE/FG03-01ER45687, the URC Summer Student
Fellowship at University of Cincinnati, and by the Chinese Academy of
sciences.  We thank Dung-Hai Lee, Pak-Wo Leung, and R. J Gooding for useful discussions.

\end{document}